\documentclass{llncs}

\usepackage{makeidx}
\usepackage{graphicx}

\newcommand{\lspm}{{LSPM}}
\newcommand{\rlspm}{{RLSPM}}
\newcommand{\gh}{{geometric hashing}}
\newcommand{\refid}{{\textit{rfid}}}
\newcommand{\grid}{{GH table}}

\newcommand{\calpha}{\ensuremath{C_{\alpha}}}

\newcommand{\mps}{\ensuremath{P_{max}}}
\newcommand{\proproThres}{\ensuremath{S_{pro}}}
\newcommand{\propatchThres}{\ensuremath{S_{patch}}}

\newcommand{\CS}{CS}

\newcommand{\eg}{{\em e.g.}}		
\newcommand{\ie}{{\em i.e.}}		

\begin{document}
\mainmatter

\title{Efficient and scalable geometric hashing method for searching
protein 3D structures}
\titlerunning{Efficient and scalable geometric hashing method for searching
protein 3D structures}
\author{
Gook-Pil Roh\inst{1} \and Seung-won Hwang\inst{1} \and Byoung-Kee
Yi\inst{2}}
\authorrunning{Gook-Pil Roh et al.}
\institute{ Department of Computer Science \& Engineering,\\
Pohang University of Science and Technology (POSTECH),\\ Pohang, Republic
of Korea\\
\{noh9pil,swhwang\}@postech.ac.kr\\
\and
Department of Medical Informatics,\\
Samsung Seoul Hospital, Republic of Korea\\
byoungkeeyi@gmail.com\\}

\toctitle{Efficient and scalable geometric hashing method for searching
protein 3D structures}
\tocauthor{Gook-Pil Roh, Seung-won Hwang, and Byoung-Kee Yi}
\maketitle

\begin{abstract}
As the structural databases continue to expand, efficient methods 
are required to search similar structures of the query structure from
the database.
There are many previous works about comparing protein 3D structures and 
scanning the database with a query structure.
However, they generally have limitations on practical use because of 
large computational and storage requirements.

We propose two new types of queries for searching similar sub-structures 
on the structural database: \lspm{} (Local Spatial Pattern Matching)
and \rlspm{} (Reverse \lspm{}).
Between two types of queries, we focus on \rlspm{} problem, because it
is more practical and general than \lspm{}.
As a na\"ive algorithm, we adopt geometric hashing techniques to 
\rlspm{} problem and then propose our proposed algorithm which improves 
the baseline algorithm to deal with large-scale data and provide an
efficient matching algorithm.
We employ the sub-sampling and Z-ordering to reduce the storage
requirement and execution time, respectively.
We conduct our experiments to show the correctness and reliability of
the proposed method.
Our experiment shows that the true positive rate is at least $0.8$ 
using the reliability measure.
\end{abstract}

\section{Introduction}
\label{sec:intro}

In this paper, we focus on searching geometrically similar proteins on
protein structural database.
Geometric similarity on protein 3D structures are known to be highly 
conserved during evolutions compared to one dimensional amino-acid
sequence identity~\cite{Orengo99From}.  
Therefore, some proteins with different sequences may indeed 
share the same or highly similar functionality, or proteins 
with high sequence homology may have different 
functionality~\cite{Whisstock03Prediction}. 
Our proposed framework, by using the protein 3D structure,
detects similar proteins with non-homologous sequences as well.

The structure-based search identifies the pairs with high structural 
similarities with the following two types:
(1) whole structure similarity comparing global structures and 
(2) sub-structure similarity comparing similar sub-structures 
shared by the pairs. 
In this paper, we focus on the sub-structure similarity, 
as it is studied to be more appropriate than whole structure 
similarity, for finding functionally related proteins~\cite{Taylor1991}.
This finding can be explained by the fact that proteins with 
similar global structures may not share similar functionality 
when their functional regions, or sub-structures, such as active or
binding sites are different~\cite{Via00Protein}.

More specifically, this paper studies the problem of searching 
similar proteins to the given query protein, which can be retrieved 
by the following two types of queries.
First, one can use a sub-structure of protein, or a patch, 
as a query and retrieve the proteins with the similar 3D structures. 
We name this type of query \lspm{} (Local Spatial Pattern Matching), 
where a query pattern is much smaller than proteins in the database.
Second, one can use a protein as a query and search the sub-structure 
databases, which we call \rlspm{} (Reverse \lspm{}). 

We aim at developing an efficient method for \rlspm{}. 
Between the two problems, \lspm{} and \rlspm{}, we focus on the latter, 
as it is more (1) practical yet (2) general.
First, regarding \emph{practicality}, while \lspm{} requires the prior 
knowledge on a query protein, by requiring users to identify ``meaningful'' 
query patches from a protein as a query.
To ensure high-quality results, those patches should be related to protein 
functions, which is hard to know in advance for a protein with an unknown 
function. 
Second, regarding \emph{generality}, \rlspm{} can be considered to be 
more general, in the sense that its solution can straightforwardly be used 
to answer \lspm{}, as the two problems are essentially the same, 
except whether query patterns are relatively larger (\rlspm{}) or 
data pattern instances are (\lspm{}).
However, while databases readily offer efficient access methods for large 
data instances, \eg, indices, it is more challenging to efficiently 
support large query structures.
In that sense, a solution to \lspm{}, not considering large queries, 
is likely to incur huge costs when applied to \rlspm{}.
However, an efficient solution to \rlspm{} designing a sophisticate 
approach to consider large query patterns will efficiently answer 
\lspm{} as well.

This paper proposes a framework for \rlspm{} based on \gh{} technique.
Our proposed framework, by identifying matches sharing functionally 
important regions, can contribute to predicting the functions of newly 
discovered proteins and moreover to designing more effective drugs.
More specifically, we summarize the contributions of our paper as follows:
\begin{itemize}
\item We develop an efficient and effective way to adopt \gh{} for our 
target problem of \rlspm{}.
\item We study an incremental maintenance scheme for grid structure 
to accommodate the insertion of large protein structures.
\item We extensively validate our proposed framework over large-scale 
data using both annotated and non-annotated proteins as queries.
\end{itemize}

The rest of the paper is organized as follows. We presents the
representation of protein 3D structure and na\"ive adoption of geometric
hashing technique to our problem in Section \ref{sec:preliminary}. Next,
we describe our proposed method in
Section \ref{sec:algorithm}. Section \ref{sec:discussion} presents our
experimental results. Section \ref{sec:conclusion} concludes our work
and presents some future work.

\section{Preliminary}
\label{sec:preliminary}

\subsection{Representation of protein 3D structure}
\label{sec:representation}
A widely adopted representation 
of the protein 3D structure
used in prior works is based on an approximation using \calpha{} and 
a pseudo atom (the centroid of the side chain).
This representation approximates all the atoms in the side chain to 
one pseudo atom to reduce both time and space complexity for protein 
comparison methods. 
However, this approximation is not appropriate of our target problem 
focusing on functional region, as atoms in side chains, among all the atoms, 
are generally associated with the functional regions~\cite{Bartlett2002}.
Therefore, approximating atoms in the side chain compromises the precision 
of the results.

In a clear contrast, we represent the 3D structures using all the atomic 
coordinates for query proteins and patches, in order not to compromise 
the precision.
As adopting this representation increases the time and space complexity, 
we will develop algorithms and access methods to optimize such cost 
as we will describe in Section~\ref{sec:algorithm}.

\subsection{Na\"ive algorithm based on geometric hashing technique}
\label{sec:gh}
This section develops a naive algorithm as a baseline approach, 
adopting \gh{} to identify functional regions of a query protein 
by scanning the patches in the database.

The \gh{} technique~\cite{Lamdan88Geom} was invented originally 
for object recognition in the field of computer vision, which has been 
adopted to compare protein 3D structures as well in many previous 
studies~\cite{Nussinov91Efficient,Fischer94Three,Wallace97TESS,XPennec07011998}.
In \cite{Nussinov91Efficient} and \cite{Fischer94Three}, they adopted 
\gh{} to identify substructures (3D motifs) of a query protein, which
are nearly identical with substructures of proteins in the database.
Pennec and Ayache proposed a comparison method for two 
given proteins. It can be extended to 
the comparison of a query protein with proteins in the
database~\cite{XPennec07011998}.
TESS~\cite{Wallace97TESS} used \gh{} for the problem similar 
to \lspm, searching proteins in the database which contain user 
defined substructure.

In contrast, our target problem is \rlspm{}, assuming 
a large-scale database of 3D substructures (patches), on which 
a whole 3D protein structure is used as a query.
While straightforwardly adopting the \gh{} technique can be 
a solution to our problem, this na\"ive approach incurs both 
space and time complexity.
In Section~\ref{sec:algorithm}, we will discuss the drawbacks of 
this na\"ive approach in details, then we will propose algorithms 
significantly outperforming it.

More specifically, the \gh{} can be adopted to our target problem 
in the following two steps--the preprocessing step and 
the matching step.
First, the \emph{preprocessing step} is applied to all patches 
to generate a \grid{} dividing a whole space into equal-sized cells 
occupied with atoms.
The origin is located at the center of \grid{} and each cell 
contains the information of atoms belonging to the cell, 
which will be used in the next step.
Then, the \emph{preprocessing step} proceeds to generate all possible 
coordinates system (\CS{}).  The \CS{} is determined by three 
non-collinear atoms, which can be one of $n(n-1)(n-2)$ possible 
combinations for a patch with $n$ atoms.
To distinguish \CS{}s, we assign each \CS{} a unique number 
(hereafter we call it \refid{}) then we store all possible 
combinations (\refid{} and coordinate) of atoms into the corresponding 
cells.

Second, the \emph{matching step} is to find similar patches 
for a given query protein. The \grid{} built on the patch database 
is matched with a set of transformed atoms by each \CS{} of 
a query protein.
As a result, we obtain a list of \refid{} pairs (\refid{} of 
the patch database, \refid{} of a query protein) which have 
matching scores larger than a user-defined threshold value.
In the matching step, a matching is carried out by searching 
the cell of \grid{} corresponding to an atom of the query protein.
If an atom $q_i$ of the query protein is located in a non-empty 
cell $C_j$ of the \grid{}, we consider that $q_i$ is \textit{matched} 
with all atoms in $C_j$.
The matching score of \refid{} pair is the number of \textit{matched} 
atoms divided by the number of atoms in the patch.
The matching step is repeated for all possible \CS{}s of the query 
protein.

\section{Method}
\label{sec:algorithm}
This section describes how to efficiently evaluate the \textsc{RLSPM}
query. We provides enhancement strategies for two steps of geometric
hashing technique. Based on these enhancement strategies, we design our
proposed algorithm. 

\subsection{Preprocessing step}
This section discusses the first step of our proposed algorithm, in which
all the atoms of each patch are transformed and inserted into a \grid{}
for all possible coordinate system.

Contrast to the na\"ive algorithm, we adopt the sub-sampling scheme
to reduce the space requirement of \grid{}.
More specifically, we define one \CS{} for each residue 
 using just three atoms (\calpha{}, N, and C atoms) of a residue.
A unique number is assigned for each \CS{}.
We call it \refid{} as in \gh{} method.
For each atom transformed by a \CS{}, we insert coordinates of the atom
and \refid{} of the CS{} into the cell of \grid{} corresponding to the coordinates
of the atom.

Recall that the na\"ive algorithm requires the space complexity $O(n^4
)$ for each protein 3D structure where $n$ is the number of atoms.
However, in the proposed preprocessing step, the total number of \CS{}
is equal to to the number of residues in protein 3D structure.
With the sub-sampling, the space complexity is reduced to $O(nm)$ where
$n$ is the number of atoms and $m$ ($<n$) is the number of residues.

In addition to sub-sampling, we also build a disk-based \grid{} 
to ensure the scalability to handle a large-scale data beyond 
the size of memory.
In a disk-based \grid{}, cells are stored in a secondary memory 
instead of a main memory. 
As a result, the size of \grid{} is not restricted within that of 
main memory.

After we insert all patches in the database into \grid{},
we then sort the cells of \grid{} in the order of z-value~\cite{oren84algo},
which enables single access per each cell during matching step.
A z-value is assigned to each cell according to its location, and 
all cells in \grid{} are sorted in the order of z-value.
The z-value is easily calculated by interleaving bit-strings of each axis-- 
When comparing two cells in two \grid{s}, we can use this unique value
to easily determine whether the two cells are in the same location.

\subsection{Matching step}
We now move on to discuss our second step of retrieving similar
patches in the database for the given query protein.
In particular, we first build a \grid{} for the given query protein, 
then match the \refid{} of query protein 
with the \refid{}s of the patches in the database.
We then retrieve all the matches with matching score higher than the given 
threshold.

As a matching score, we use the ratio of the number of overlapped cells 
in a query protein with the cells in a patch to the number of cells in 
a patch.  
Note our matching score is defined in the level of cells and not 
in the level of atoms.
However, we can adjust the cell size $\delta$, to decrease to obtain 
the results with high accuracy or to increase for efficiency.

With the matching score metrics, we now describe the overall 
structure of our algorithms.
First, a \grid{} is built for a query protein with the similar way of
the preprocessing step.
The only difference from building the \grid{} for patch database is that 
we only consider atoms within \mps{} from origin.
We set \mps{} to the size of the largest patch, the size of matches is 
always less than \mps{}.
Hereafter we call the \grid{} of a query protein $G_q$ and the \grid of 
patch database $G_p$.
Second, for each \grid{}, the cell with the smallest z-value is loaded 
into the main memory. As the \grid{} was sorted in the order of z-value, 
this is done by scanning each \grid{} in the sorted order.
During this scan, z-values of current two cells are checked-- If the 
z-value of a cell from $G_p$ is bigger than that from $G_q$, the next cell 
of $G_q$ is fetched into the memory and the equality test is performed 
again for a newly loaded cell.

When the two cells have the same z-values, we update the matching scores 
between all \refid{} pairs of a patch and a query protein.
Update can be done by  adding the number of atoms with the same \refid{} 
in a cell of $G_p$ to the previous score of each \refid{} of a query protein. 
The matching scores are stored in a temporary file 
(hereafter called the \emph{score table}) in case its size is too large to keep 
in the main memory.

After the above update processes for all the overlapped cells, all the values 
in the score table, \ie, \refid{} of patch database, \refid{} of a query 
protein, is divided by the size of the patch corresponding to \refid{} of 
patch database. The size of patch is the number of all atoms in it. 
From the score table, we extract only the \refid{} pairs with matching 
scores higher than the given threshold value called protein-patch 
threshold \propatchThres{}. 
In other words, our proposed algorithm will report the sub-regions of 
a query protein which are structurally similar to some patches with 
higher matching scores than \propatchThres{}.

\section{Experimental results}
\label{sec:discussion}
In this section, we first describe how to build the patch database
from existing databases 
such as PDB~\cite{Berman00PDB} and CSA~\cite{catalytic2004Craig} 
in Section~\ref{sec:ppd}.
We then validate the reliability of our proposed algorithm over all
protein 3D structures from PDB.
More specifically,  
for the proteins without annotated patches, we compute reliability using 
`keyword recovery' used in the field of protein-protein interaction 
(Section~\ref{sec:noannoprotein}).

\subsection{ Protein Patch Database }
\label{sec:ppd}
In this section, we discuss how we build PPD (Protein Patch Database), 
by extracting functional regions (patches) from two existing protein 
structure databases, \ie, PDB and CSA.

First, PDB~\cite{Berman00PDB} contains the 3D coordinates of atoms 
and functional information. 
For some proteins in this database with ``known" functional regions, 
residues directly involved with those regions are annotated (in `SITE' record).
We collected atomic coordinates for those residues, \ie, $9206$ patches, 
inserted into PPD.

Second, CSA~\cite{catalytic2004Craig} contains enzyme active sites and 
catalytic residues in enzymes, based on the functional annotations of 
PDB and SWISS-PROT database.
More specifically, CSA provides two types of structural templates--
Ca/Cb atom template and functional atom template.
A Ca/Cb atom template provide only Ca or Cb atom coordinates and a functional 
atom template provide atomic coordinates of directly related atoms 
to interact its substrates.
In particular, we use 
the Ca/Cb atom template of CSA version $1.0$ to obtain functionally 
important residues.
For these residues, we then generate a patch, by extracting atomic 
coordinates belong to those residues, \ie, $147$ templates in 
CSA version $1.0$. Among them $34$ templates were identical to the patches 
extracted from PDB as CSA is based on PDB.
We remove such duplicates and insert the remaining $113$ patches of 
CSA into PPD.

\newcommand{\pid}{1H75}
\newcommand{\patch}{\pid\_0}
\newcommand{\qid}{1TDE}

\subsection { Reliability analysis against non-annotated protein}
\label{sec:noannoprotein}
In this section, we discuss our validation using non-annotated 
proteins.
Non-annotated proteins mean that there is no patch annotated on them.
Therefore, when using the non-annotated proteins as queries, the matching
results are expected to contain matching pairs of a query protein and
matching patches that are not annotated on the query protein.

We first discuss the measure that can be used to analyze the reliability 
of the results when using non-annotated proteins as queries.
We will then show our reliability results.

As a reliability metric, we adopt `keyword recovery'~\cite{sprinzak2003} 
used in the field of PPI (Protein-Protein Interaction) for validating results.
More specifically, this metric compares the annotation keywords of 
two proteins and calculates the true positive (TP) rate. 
As performing \textit{in vivo} or \textit{in vitro} experiments to validate 
the functional relationships of every pair is infeasible, we adopt this 
indirect measure instead.

The TP rate is calculated by following formula~\cite{sprinzak2003}.
\[ TP = (D-R)/(I-R) \]
where $D$ is the ratio of pairs with the same keywords from the given 
dataset. $R$ is the ratio of pairs with the same keywords among all pairs 
in the given dataset.
$I$ is the ratio of pairs with the same keywords in true matching pairs. 
We set $I$ to $1$ as in~\cite{sprinzak2003}.

To adopt the `keyword recovery' method, annotated keywords are needed 
for query proteins and patches. We use keywords about the biological 
process and cellular component in Gene Ontology~\cite{GO}.
In case of the patch, we use keywords of the protein from which the patch 
has been extracted, because there is no annotation for a particular patch.

Using the reliability measure, we calculate the true positive rate for
the matching result which is a list of matching pairs with higher
matching scores than \propatchThres{}.
In addition to \propatchThres{}, we introduce an another parameter
\proproThres{} that restricts matching pairs whose structural identity
is bigger then \proproThres{} to eliminate the redundancy in the
matching result. 
We perform our algorithm to compute structural identity between the
query protein and the protein from which ``matched'' patch is extracted.
If the score between two proteins is bigger than some threshold value 
(\proproThres{}), 
two proteins are regarded as same protein and corresponding matching pair 
is thrown away from the result set for validation. 

\begin{figure}[ht]
\centering
\includegraphics[width=8.6cm]{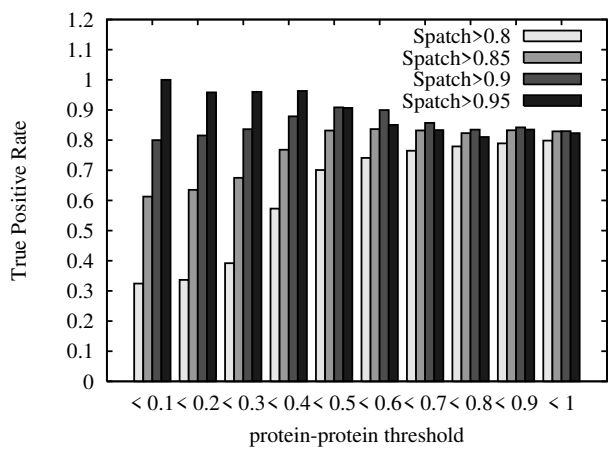}
\caption{True Positive rate by various threshold values}\label{fig:tprate}
\end{figure}

Figure~\ref{fig:tprate} shows the TP rates for varying \propatchThres{} and \proproThres{}.
We only present the results when the biological process is used as keywords. Because there are few entries on which cellular components are annotated, we omit the result of cellular component.
We consider \propatchThres{} to be in the range from $0.8$ to $0.95$.
When \propatchThres{} is low, it frequently occurs for a query protein to be matched with a patch by chance.
This means lots of meaningless matching pairs could be included in the result. 
Therefore, We remove those matching pairs from the result by setting \propatchThres{} to a value higher than $0.8$.  

As \propatchThres{} and \proproThres{} is set to $0.8\sim0.95$ and
$0.1\sim1$ respectively, we calculate the TP rates of the results given
by two threshold values.
As shown in Figure~\ref{fig:tprate} The TP rates decrease for lower value of \propatchThres{} when \proproThres{} is less than $0.5$.
This observation indicates that the number of pairs (query protein, patch) which are matched by chance increases according as \propatchThres{} is decreasing.
As \propatchThres{} increases (i.e. more precise matching) overall TP
rate also increases. This shows the specificity of functional site. 

Two parameters, \propatchThres{} and \proproThres{} have influence on
the reliability of the matching result.
We recommend that \propatchThres{} is set to the value between $0.9$ and $1$. Because the interaction between proteins generally has high specificity, the structural similarity between a query protein and a patch should be high to obtain the meaningful matching results.
For \proproThres{}, we recommend users to set it according to the application.
Our experiment shows that the TP rate is at least $0.8$ when \proproThres{} and \propatchThres{} are set as our recommendation.

\section{Conclusion}
\label{sec:conclusion}
In this paper, we study sub-structure similarity search on protein 3D
structural database. We present two types of sub-structure similarity search:
LSPM and reverse LSPM. Between them, we focus on reverse LSPM problem
because it is more practical and general then LSPM.
Toward the goal, 
we introduced our improved algorithm 
which significantly outperforms adopting geometric hashing technique
``as is" in terms of both storage overhead and execution time.
More specifically, to reduce storage overhead, we applied a sub-sampling 
technique to coordinate system set and developed a disk-based \grid{} for accommodating 
a large scale of query protein patches in the database.
Furthermore, to reduce execution time, we restricted a query structure 
within a reasonable range, namely within maximum patch size, and employed 
the Z-ordering to eliminate redundant accesses which
requires only a single access per a cell of \grid{} by concurrently
scanning two \grid{} sorted in the order of z-values.

The reliability of proposed method was validated using our
protein patch database which we build by extracting annotated residues
from PDB and CSA. 
The true positive rate is at least $0.8$ under
recommended parameter value. We are in the process of validating the
reliability of our method over other protein structural database such as
protein-protein interface database~\cite{scoppi}.

\bibliographystyle{plain}

\end{document}